\documentclass[
 reprint,
 amsmath,amssymb,
 aps,prb,eps,superscriptaddress,twocolumn,10pt
]{revtex4-1}

\usepackage{graphicx}% Include figure files
\usepackage{dcolumn}% Align table columns on decimal point
\usepackage{bm}% bold math
\usepackage{xcolor}
\usepackage{hyperref}

\begin{document}

\preprint{APS/123-QED}

%\title{Role of correlations in the estimation of Curie temperature for CoBr$_2$, a possible 2D ferromagnet}
%\title{Theoretical evidence for ferromagnetism in the two-dimensional material CoBr$_2$}
\title{Importance of electronic correlations for the magnetic properties of the two-dimensional ferromagnet CoBr$_2$}

\author{Hrishit Banerjee}
 \email{h.banerjee10@gmail.com}
\author{Markus Aichhorn}%
\affiliation{Institute of Theoretical and Computational Physics, Graz University of Technology, NAWI Graz, Petersga{\ss}e 16, Graz, 8010, Austria.
}

\date{\today}% It is always \today, today,

\begin{abstract}
We investigate the emergence of ferromagnetism in the two-dimensional metal-halide CoBr$_2$, with a special focus on the role of electronic correlations. The calculated phonon spectrum shows that the system is thermodynamically stable unlike other Co halides. We apply two well-known methods for the estimation of the Curie temperature. First, we do DFT+U calculations to calculate exchange couplings, which are subsequently used in a classical Monte Carlo simulation of the resulting Ising spin model. The transition temperature calculated in this way is in the order of 100\,K, but shows a strong dependence on the choice of interaction parameters.
Second, we apply dynamical mean-field theory to calculate the correlated electronic structure and estimate the transition temperature. This results in a similar estimate for a noticeable transition temperature of approximately $100$\,K, however, without the strong dependence on the interaction parameters. 
\textcolor{black}{The effect of electron-electron interactions are strongly orbital selective, with only moderate correlations in the three low-lying orbitals (one doublet plus one singlet), and strong correlations in the doublet at higher energy. This can be traced back to the electronic occupation in DMFT, with five electrons in the three low-lying orbitals and two electrons in the high-energy doublet, making the latter one half-filled. Nevertheless, the overall spectral gap is governed by the small gap originating from the low-lying doublet+singlet orbitals, which changes very weakly with interaction $U$. In that sense, the system is close to a Mott metal-to-insulator transition, which has been shown previously to be a hot-spot for strong magnetism.}
%We attribute the rather high transition temperature to the vicinity of the system to a paramagnetic Mott metal-to-insulator transition\textcolor{black}{, in the sense that the correlated spectral function is insulating with a very small spectral gap. The effect of correlations are strongly orbital selective, with } The robustness of the transition temperature with respect to the interaction values can be traced back to the special electronic configuration with five electrons in the \textcolor{black}{three low lying orbitals (one doublet+one singlet) which are almost degenerate and two electrons in the two higher energy doublet degenerate orbitals}, showing a  very weak dependence of the single-particle gap on the Hubbard interaction $U$.
%
%We study the effect of correlations in predicting the Curie temperature for CoBr$_2$, touted as a possible 2D ferromagnet with a reasonably low exfoliation energy, from a first principles perspective. We show that correct description of Hubbard U is extremely important in case of DFT+U methods to obtain the correct magnetic exchange, which contributes heavily to Curie temperature in the standard methods of determining the Curie temperature by solving Ising models with real spin quantum numbers using classical Monte Carlo methods. However since the material is a small band gap insulator in the paramagnetic phase, DMFT does a better job of describing the correct Curie temperature, independent of Hubbard U and J values. 
\end{abstract}

\maketitle

\section{Introduction}
There has been a lot of recent excitement about functional two-dimensional (2D) materials, which  
 provide opportunities to venture into largely unexplored regions of materials space. On one hand, their thin-film like nature makes them extremely promising for applications in electronics. On the other hand, the physical properties of monolayers often differ dramatically from those of their parent three-dimensional materials, providing a new degree of freedom for applications while also unveiling novel physics associated with low dimensionality. Moreover, van-der-Waals (vdW) heterostructures have recently emerged as an additional avenue to engineer new properties by stacking 2D materials in a desired fashion.

Emergence of spontaneous ferromagnetism (FM) without doping in 2D materials has been receiving a lot of attention, since long-range FM in 2D can facilitate various applications.\cite{Gong2017, Huang2017, Mak2019} According to the Mermin-Wagner theorem,\cite{Mermin-Wagner} continuous symmetries cannot be spontaneously broken at finite temperatures in systems with sufficiently short-range interactions in dimensions $D\leq 2$. This implies that ferromagnetism cannot be stabilised in 2D without additional symmetry-breaking effects. The additional symmetry breaking may be provided by the presence of sufficiently strong spin-orbit coupling (SOC), resulting in magnetic anisotropy. This requirement in low-dimensional systems therefore explains the rareness of inherent 2D FM materials. Such anisotropic symmetry breaking has recently been observed in monolayers of CrI$_3$ and Cr$_2$Ge$_2$Te$_6$, leading to spontaneous stable ferromagnetism.\cite{Klein1218, Jiang2018, Torelli_2018, Xu2018, WANG2019293}  
These studies have shown the emergence of spontaneous magnetism in 2D originating from the transition metal $d$ orbitals. These materials are insulating with small band gaps. 

There have been several first-principle predictions of the ferromagnetic Curie temperature $T_C$ in 2D materials in general. Most of these predictions follow the well-known procedure of solving an Ising or Heisenberg model using Monte Carlo methods,\cite{Han2020, Kabiraj2020, chen-nano-2020, Lu2019, WANG2019, Miao} where the magnetic superexchange parameters for those models are extracted from density-functional theory (DFT)+U calculations. A recent development in this field is to use high-throughput machine learning methods
%, based on DFT+U based Monte Carlo studies to to train their data sets and 
to estimate transition temperatures for certain materials.\cite{Kabiraj2020} 
Notwithstanding the fact that these Monte-Carlo methods tend to overestimate the Curie temperatures by some amount, there is also the additional problem of how to correctly determine the magnetic exchange coupling and magnetic anisotropy from DFT+U calculations. They depend heavily on the choice of Hubbard $U$ and Hund $J_H$ parameters, particularly in these strongly-correlated $d$-shell transition metals in which such spontaneous magnetism is seen.

A recent high-throughput study predicted the possibility of exfoliation of monolayers from a significant number of experimentally available materials,\cite{Mounet2018} which may show intrinsic ferromagnetism in its monolayer form. A significant class of materials among them belong to the MX$_2$ class of metal halides. Since metal halides are van-der-Waals crystals, they have low exfoliation energy in general. In addition, the associated magnetic anisotropy makes them ideal candidates for the emergence of intrinsic 2D ferromagnetism. In this study, we concentrate on one member of the CoX$_2$ class of materials. %, which is ripe for correlations based study.
Since CoCl$_2$ and CoI$_2$ are possibly structurally unstable, as it is seen from negative frequencies in the phonon excitation spectrum.\cite{Mounet2018}
We thus focus on CoBr$_2$ and study this particular material in detail, primarily from the point of view of strong electronic correlations. Needless to say that such methodology may be applied to other relevant metal halides or 2D materials in general as well.

Several interesting properties for CoBr$_2$ have been found in experimental studies, as well as predicted from first-principle calculations. In a theoretical work, a topologically nontrivial insulator state with a quantum anomalous Hall effect and a topological Chern number $Z=4$ has been predicted, and it has been shown that its edge states can be manipulated by changing the width of its nanoribbons and by applying strain.\cite{Chen2017} Very recently it has been shown that biaxial tensile strain can induce an FM to antiferromagnetic (AFM) phase transition in the CoBr$_2$ monolayer, while compressive strain stabilises the ferromagnetic ground state. Furthermore, doping plays obviously a critical role in changing the ground state from a semiconductor to a half metal, which is particularly important for spintronics based applications.\cite{Sun2019} The same study also hinted at a possibly large $T_C$. However, a recent study on metal halides predicts a small Curie temperature of 24\,K, albeit with a large magnetic exchange of 6.7\,K.\cite{Botana2019} 
The large dependence of the Curie temperature on the actual choices for the parameters Hubbard $U$ and Hund exchange $J_H$ has been noted, as well as a moderate $T_C \sim 0.94 \times T_C^{CrI_3}  \sim 43K$ has been estimated in another study.\cite{Liu2018} Thus, it is imperative to understand the electronic, and in particular magnetic properties of the CoBr$_2$ monolayer better. Since Co happens to be a strongly-correlated $d$ shell transition metal ion, this has to be done with a special focus on the description of electronic correlations.

In this work we apply two well established methods for the estimation of the transition temperature of CoBr$_2$. First, we calculate exchange couplings and the magnetic anisotropy using DFT+U methods. We find a strong variation of these couplings on the parameters $U$ and $J_H$, which in turn influences strongly the predicted Curie temperature from a classical Ising model Monte Carlo simulation.
%in the standard methods of determining the Curie temperature by solving Ising models with real spin quantum numbers using classical Monte Carlo methods. 
%The value of magnetic superexchange varies significantly with change in Hubbard U with fixed Hund exchange $J_H$, and this causes significant change in the Curie temperature calculated using classical Markov chain Monte Carlo methods.  
Second, we apply the dynamical mean-field theory (DMFT) to the problem, and calculate the magnetisation as function of temperature in order to estimate $T_C$. We show that the material in its paramagnetic state \textcolor{black}{has a very small, almost vanishing total spectral gap, changing only slightly with interaction parameters. In that sense, we call the system close to a metal-insulator phase transition. When looking at the electronic correlations in an orbital-resolved manner, however, one can see strong orbital selectivity of the correlation effects. DMFT polarises the electronic occupancies of the $d$-orbitals, making the higher-energy doublet orbitals half-filled and very susceptible to correlations, whereas the gap in the lower-energy orbitals is small and shows almost no dependence on the interaction parameters. The vicinity to the metal-to-insulator phase transition has been argued in previous works to be highly beneficial for magnetism,\cite{onebanddmft,jernej, alen2017} and we argue that this mechanism is also at work here.}
%we attribute the rather high transition temperatures found here to this position of the 
%As a result, the estimated $T_C$ is quite independent on the choice of parameters.

\section{Computational Details}
Our DFT calculations for structural relaxation were carried out in a plane-wave basis with projector-augmented wave (PAW) potentials~\cite{blochl} as implemented in the Vienna Ab-initio Simulation Package (VASP).\cite{kresse, kresse01}
For our DFT+DMFT calculations we are using the full-potential augmented plane-wave basis as implemented in the \textsc{wien2k} code package.\cite{wien2k}

In all our DFT calculations, we chose as exchange-correlation functional the generalized gradient approximation (GGA), implemented following the Perdew Burke Ernzerhof (PBE) prescription.\cite{pbe} 
The DFT+U calculations were carried out in the form of GGA+U. The value of $U$ at the Co sites in the GGA+U scheme was varied between 3.5 and 4.5\,eV, with a fixed Hund's exchange $J_H$ of 1\,eV. 
We note here that the choice of Hubbard $U$ and Hund $J_H$ was inspired by the choices of $U$ and $J_H$ in recent literature.\cite{Botana2019, Sun2019, Liu2018} It is also seen in general in DFT calculations, a slightly larger $J_H \sim 1$ favours ferromagnetism.\cite{millis} %Our determination of the magnetic ground state involves comparison of FM and AFM energies in VASP.

For ionic relaxations using the VASP package, internal positions of the atoms were allowed to relax until the forces became less than 0.005\,eV/\AA. An energy cutoff of 550\,eV and a 6$\times$12$\times$4 Monkhorst–Pack $k$-points mesh provided good convergence of the total energy. Spin-orbit coupling was taken into account in a perturbative non-self-consistent manner as implemented in VASP. A vacuum thickness of about 15\,$\AA$ was found to be sufficient to get rid of any spurious electric field effects. The phonon spectrum was calculated based on the density functional perturbation theory (DFPT) as implemented in the VASP package. A 3$\times$3$\times$1 supercell and a $\Gamma$-centered 3$\times$3$\times$1 Monkhorst-Pack $k$-point mesh  were  used.  The  phonon  frequencies were calculated using the Phonopy code.\cite{phonopy}

For the \textsc{wien2k} calculations, we used the largest possible muffin-tin radii, and the
basis set plane-wave cutoff was defined by ${R_{\text{min}}\!\cdot\!K_{\text{max}}} = 7.5$, where $R_{\text{min}}$ is the
muffin-tin radius of the Br atoms. The consistency between the VASP and \textsc{wien2k} results has always been cross-checked. 
We perform the DMFT calculations in a basis set of projective Wannier functions, which were calculated using the DFTTools package\cite{aichhorn1, aichhorn2, aichhorn3} based on the TRIQS libraries.\cite{triqs} For our calculations, all five Co $d$ orbitals have been taken into account in the correlated subspace. The Anderson impurity problems were solved using the continuous-time quantum Monte Carlo algorithm in the hybridization expansion (CT-HYB)\cite{werner06} as implemented in the TRIQS/CTHYB package.\cite{pseth-cpc} We performed one-shot DFT+DMFT calculations, with an FLL-type double counting correction as given in Ref.~\onlinecite{helddc} 
%Held type double-counting correction~\cite{anisimov93}.
We use a fully rotationally-invariant Kanamori Hamiltonian parametrised by Hubbard $U$ and Hunds coupling $J_H$, where we set the intraorbital interaction to
$U'=U-2J_H$.
%both the density-density as well as rotationally-invariant Kanamori interaction~\cite{kanamori63}. 
For our DMFT calculations we used $U$ values ranging from 3.5 to 4.5\,eV and $J_H = 0.5$\,eV in order to investigate the effect of the interaction parameters on $T_C$. The choice of interaction parameters is motivated by previous DMFT work on cobalt-oxide compounds such as Na$_x$CoO$_2$, showing a very similar layered crystal structure. There, excellent agreement of DMFT with experimental ARPES band structure as well as prediction of experimental properties driven by correlations have been demonstrated for a similar range of $U$ and $J_H$ values.\cite{parcollet, held}

We note here that there is however no reason for the two sets of Hubbard $U$ and Hunds exchange $J_H$ parameters between DFT+U and DMFT schemes to be identical. This is simply due to the reason that the local orbitals  implemented within VASP are quite different from the low-energy Wannier projections used in DMFT, with the Wannier orbitals being more extended in space. In general a slightly smaller value for $U$ and $J_H$ is expected to be required to correctly estimate the electronic properties in case of a DMFT calculation compared to the DFT+U methods.

Real-frequency spectra have been obtained using the maximum-entropy method of analytic continuation as implemented in the TRIQS/MAXENT application.\cite{maxent}

\section{Results}
\begin{figure}[tb]
    \centering
    \includegraphics[width=0.9\columnwidth]{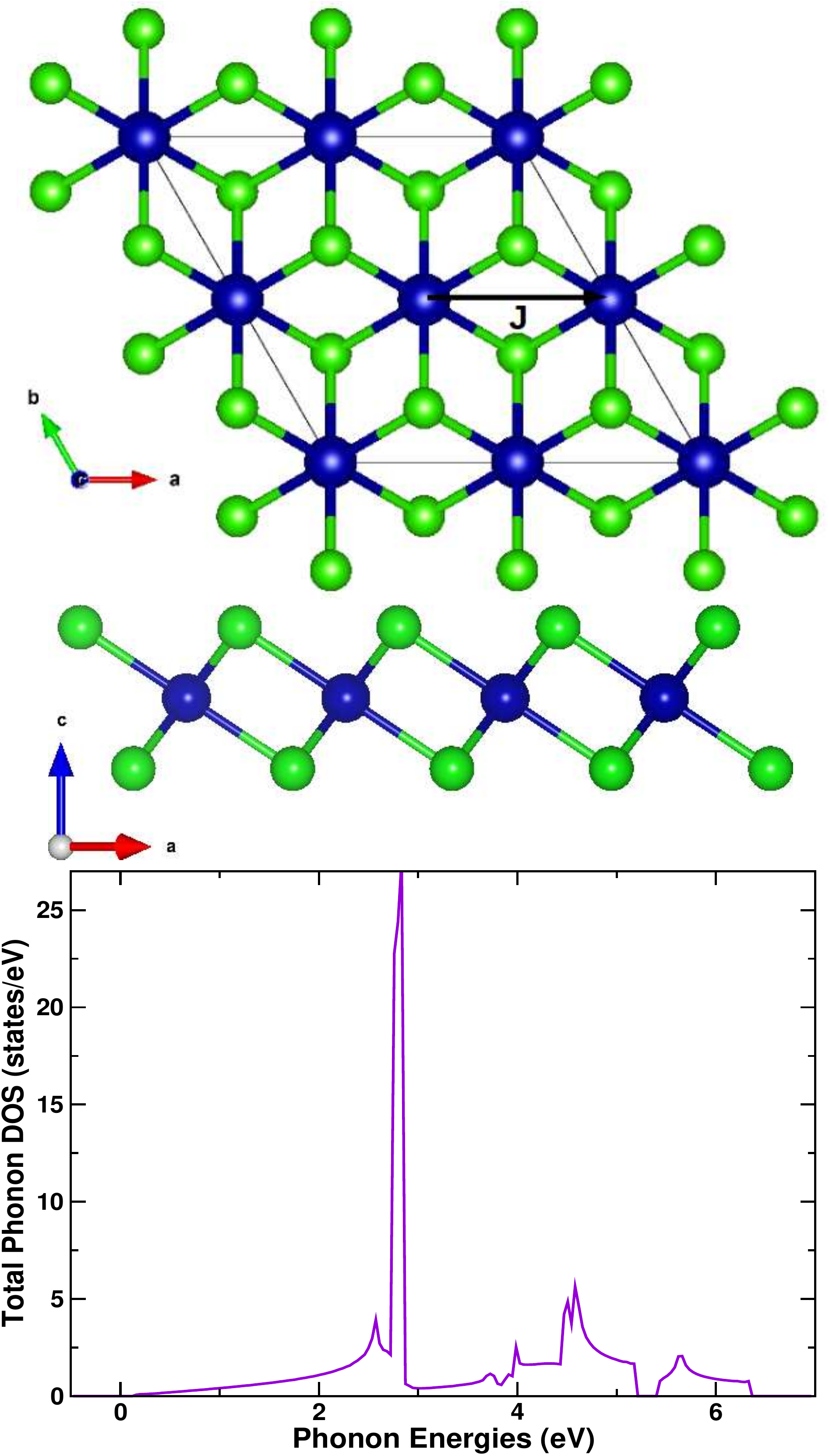}
    \caption{(Color online) Top: Crystal structure of a monolayer of CoBr$_2$. Co atoms are shown in blue, and the surrounding Br atoms in green. \textcolor{black}{Middle: Crystal structure showing a lateral image of the monolayer to highlight the buckling of the structure.} The magnetic superexchange coupling $J$ is also indicated as black line. Bottom: Total phonon density of states. The absence of spectral weight at negative frequencies confirms the structural stability of the material.}
    \label{fig1}
\end{figure}

\subsection{Crystal and DFT electronic structure}

First we describe the crystal structure of CoBr$_2$. It is a van-der-Waals crystal with symmetry P$\bar{3}m_1$ and lattice constants $a=b=3.738$\,\AA, $c=16.907$\,\AA{} and $\alpha=\beta=90^o$, $\gamma=120^o$. It has been shown\cite{Mounet2018} that it has a low exfoliation energy of 16.8 meV/$\AA^2$. It is a buckled rather than a planar structure, with Co-Br-Co out of plane angles of 92.3$^o$. Each Co has 6 Br nearest neighbours which form magnetic superexchange paths to other Co atoms. The structure of a monolayer of CoBr$_2$ along with the magnetic superexchange path $J$ is shown in the upper panel of Fig.~\ref{fig1}.   

To determine the dynamical stability of the CoBr$_2$ monolayer we first relax the ionic positions, and then carry out phonon calculations for the relaxed structure within VASP, using a 3$\times$3$\times$1 supercell. The total phonon density of states (DOS) is shown in the lower panel of Fig.~\ref{fig1}. Unlike CoCl$_2$ and CoI$_2$,\cite{Mounet2018} we do not see any negative frequencies for CoBr$_2$. Thus we can ascertain the dynamical stability of the CoBr$_2$ monolayer.

\begin{figure}[tb]
    \centering
    \includegraphics[width=0.9\columnwidth]{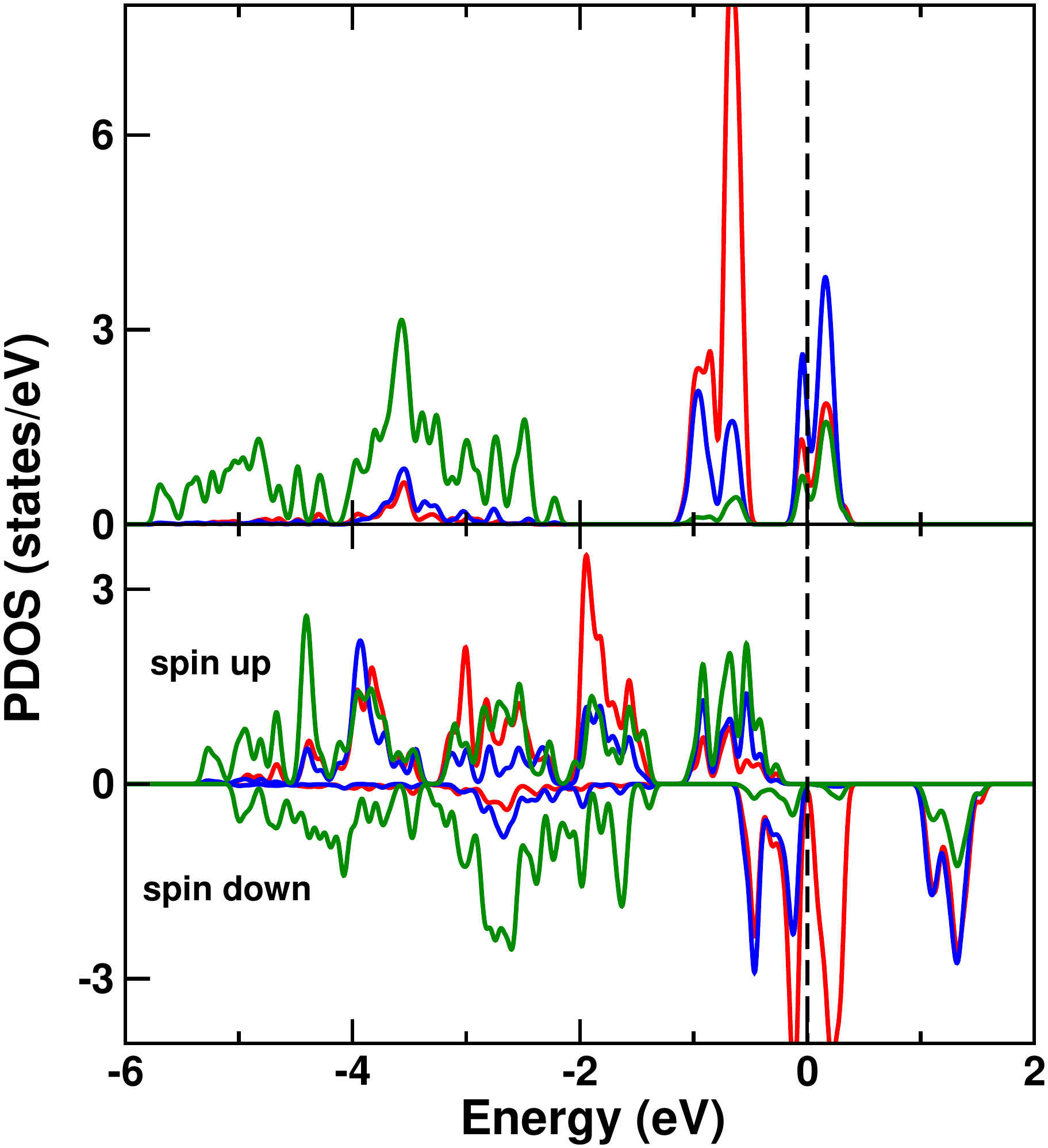}
    \caption{(Color online) DFT electronic structure of a monolayer of CoBr$_2$. The top panel shows the non-spin-polarised DOS while the bottom panel shows the spin-polarised DOS, both projected to the relevant orbitals. \textcolor{black}{The green lines represent the Br $p$ orbitals, the red lines represent the sum of the lower energy singlet+doublet $d$ orbitals which are almost degenerate, and the blue lines represent the higher energy doublet.}}
    \label{fig2}
\end{figure}
In Fig.~\ref{fig2} we show the PBE electronic structure for the monolayer, both in the spin-polarised and non-spin-polarised ground states. In the top panel of Fig.~\ref{fig2} we show the non-spin-polarised projected density of states (PDOS), projected to the relevant Co $d$ orbitals. We see a metallic ground state with a mix of Co \textcolor{black}{$d$} orbitals at and around the Fermi energy, which is marked by the dashed line, with some mixing from Br $p$ orbitals. \textcolor{black}{A combination of almost degenerate orbitals mostly lie at a lower binding energy of -0.8\,eV} with respect to Fermi energy, while \textcolor{black}{higher energy doubly-degenerate} orbitals occupy the states at the Fermi energy. \textcolor{black}{We explain this arrangement of $d$ orbitals in more detail in Appendix A and show the band structure along with the corresponding Wannier projections in Fig. \ref{figA1}. A diagonalization of the local Wannier Hamiltonian shows that the three lower energy states are split up into a singlet at -0.849\,eV, and a doublet very close in energy at -0.842\,eV, resp. The two degenerate orbitals around the Fermi level are found at 0.043\,eV, forming the higher-energy doublet. At the level of non-magnetic DFT calculations, all the three lower energy singlet+doublet orbitals have almost the same filling in Wannier space, summing up to $n=5.85$ electrons, and the higher energy doublet is occupied by $n=1.15$ electrons.} The energies below -2\,eV are dominated by the Br $p$ orbitals. The spin-polarised PDOS in the bottom panel of the same figure shows a half-metallic state with a large spin splitting and a calculated moment of 3\,$\mu_B$. An ad-hoc application of static Hubbard $U$ as done in general in DFT+U calculations gives a very large band gap.
%, for a material at the metal insulator transition point. 
We do not show here the PDOS from our PBE+U calculations, but refer to a recent study which shows the this band structure, and also confirms the phonon bands for \textcolor{black}{this system}.\cite{Sun2019}

\subsection{DFT+U combined with Monte Carlo study}
\begin{figure}
    \centering
    \includegraphics[width=0.95\columnwidth]{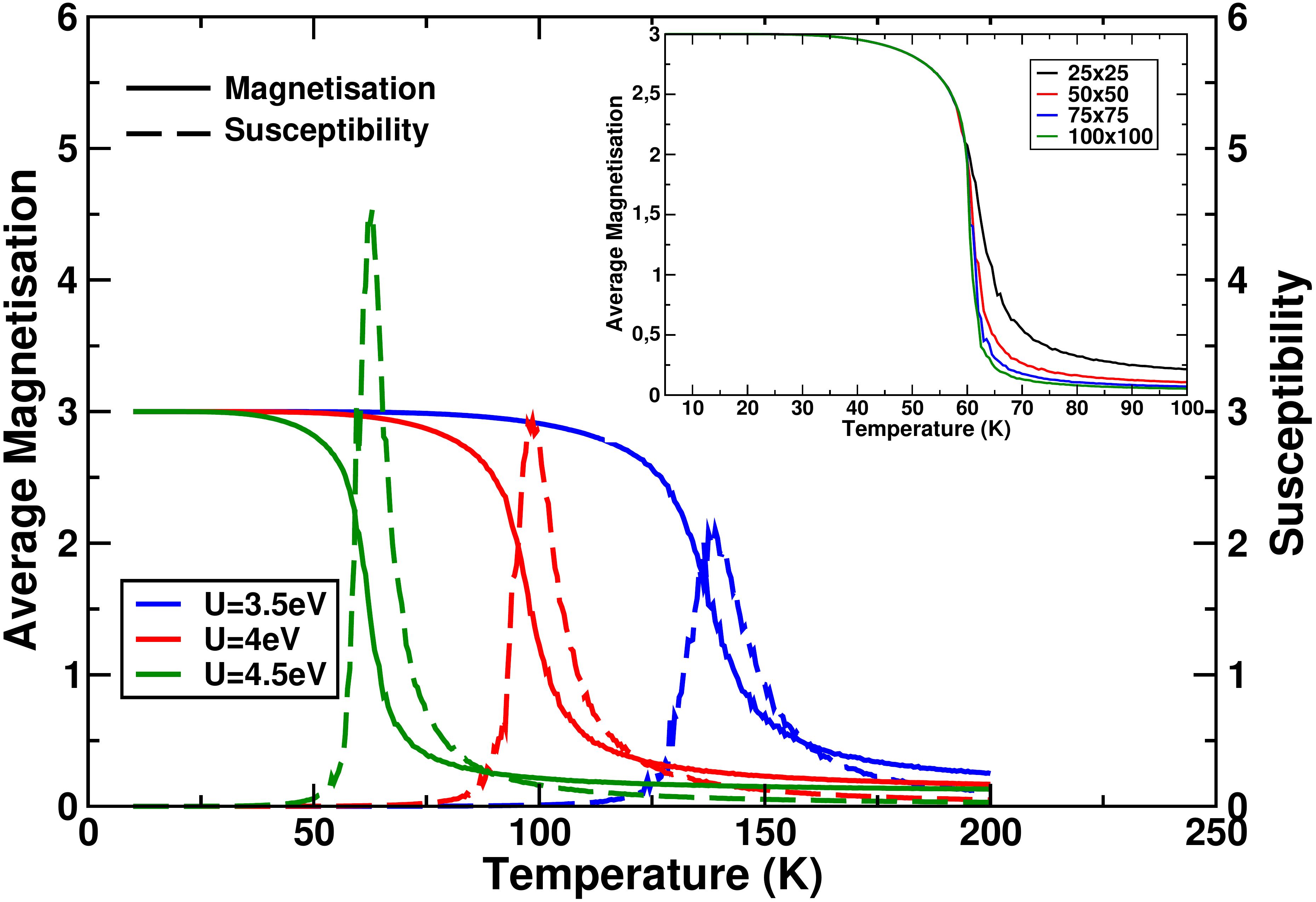}
    \caption{(Color online) Magnetisation and susceptibility obtained from Monte Carlo simulations of a Ising spin model \textcolor{black}{on a 25$\times$25 sites lattice}. The magnetic superexchange $J$ and magnetic anisotropy energy $A_M$ are extracted for three different values of Hubbard $U=3.5$\,eV, 4\,eV, and 4.5\,eV. \textcolor{black}{Inset: Effect of the finite cluster size on the average magnetisation, shown for lattice sizes of 25$\times$25, 50$\times$50, 75$\times$75, 100$\times$100, for the case of $U=4.5$\,eV.}}
    \label{figmagn}
\end{figure}

We next determine the magnetic superexchange as well as the magnetic anisotropy energy from first principles. For calculating the magnetic exchange coupling $J$, a 2$\times$1$\times$1 supercell was constructed and internal positions were relaxed. Next, self-consistent energy calculations for $U=3.5$\,eV, 4\,eV, and 4.5\,eV (for a fixed $J_H=1$\,eV) for both ferromagnetic and anti-ferromagnetic configurations were carried out. The total energies from these calculations were fitted to a simple \textcolor{black}{nearest-neighbor} Ising model \textcolor{black}{$H=-J \sum_{\langle ij\rangle}S^z_iS^z_j$} to obtain the coupling $J$. The magnetic anisotropy energy $A_M = E_{SOC}^{z}-E_{SOC}^{xy}$ was calculated as the difference in energy when the magnetic moment is pointing in the $z$ direction or in the $xy$ plane, resp. Of course, spin-orbit coupling is necessary to be included in these calculations.

From our DFT+U calculations we see that $J$ varies significantly with changes in the Hubbard parameter $U$ for fixed $J_H=1$\,eV, and this variation is shown in Table \ref{tab1}. The anisotropy $A_M$ is calculated to be $-0.4$\,K. A magnetic superexchange value similar to our $J$ for Hubbard $U=3.5$\,eV has been calculated in a recent study,\cite{Botana2019} however, the explicit variation with $U$ and $J_H$ has not been discussed.

The Ising model is then constructed for a 2D lattice according to the equation 
\textcolor{black}{$$ H=-J \sum_{\langle ij\rangle} S^z_iS^z_{j}+A_M \sum_{i}S^z_iS^z_i $$}. The magnetisation and the susceptibility of this model are then calculated with a Markov chain Monte Carlo algorithm using the Metropolis-Hastings rejection scheme. 
\textcolor{black}{All results shown here are calculated on the 2D triangular lattice of Co atoms (see Fig.~\ref{fig1}) %using a total of $25^2$ lattice sites 
with periodic boundary conditions. }
%For taking into account the exchange interactions on a triangular lattice within the scheme of a square array, we consider the exchange interactions between the site $(i,j)$ with both sites $(i, j+1)$ and $(i+1,j+1)$, with periodic boundary conditions. Thus all six possible nearest neighbor exchange interactions are accounted for.
%The finite size scaling effect in the Monte Carlo simulation results has been checked and a system containing 25$\times$25 lattice sites, has been found to be large enough to avoid finite size effects.
\begin{table}[]
    \centering
\begin{tabular}{|c|c|c|}
	\hline 
$U$ (for $J_H=1$\,eV)	& $J/S^2$ (K) & $T_C$ (K) \\ 
	\hline 
3.5\,eV	& 6.4 & 139 \\ 
	\hline 
4.0\,eV	& 4.5 & 99 \\ 
	\hline 
4.5\,eV	& 2.8 & 63 \\ 
	\hline 
\end{tabular} 
\caption{Variation of coupling $J$ and transition temperature $T_C$ with changing Hubbard parameter $U$.}
    \label{tab1}
\end{table}

We show in Fig.~\ref{figmagn} a plot of the magnetisations and susceptibilities calculated from Monte Carlo simulations \textcolor{black}{on a 25$\times$25 lattice}. We see, for all values of $U$, a transition from a high-spin ferromagnetic state with magnetic moment of 3\,$\mu_B$ to a paramagnetic state without a net magnetic moment. The transition temperature, however, depends quite significantly on the value of $U$. The peak in the susceptibilities shows the phase transition point. For the three different $U$ values of 3.5\,eV, 4\,eV, and 4.5\,eV the model predicts transition temperatures of 139\,K, 99\,K and 63\,K respectively. All these values of $T_C$ are larger compared to the experimentally measured $T_C$ of 45\,K for CrI$_3$. 
%albeit being overestimated due to the use of a Monte Carlo Ising model. 
Irrespective of a possible overestimation of the transition temperature calculated from Monte Carlo, we note here a large variation of almost $\sim 40$\,K with change in Hubbard $U$ parameter of 0.5\,eV in each case. This dependence is what we intend to address in our next section with an investigation using DMFT.

\textcolor{black}{We have checked the results for larger lattice sizes of 50$\times$50, 75$\times$75, and 100$\times$100, and we do not see any appreciable change in the $T_C$, see the inset of Fig.~\ref{figmagn}. The magnetisation curves become sharper indicating a sharper transition at the larger size of the lattice. Putting $A_M=0$\,K, we find a significant shift of the transition to the left in all the curves (not shown). However, the transition does not vanish completely, simply due to the use of an Ising model, which has an inherent anisotropy and hence supports a magnetic transition in two dimensions. Our goal here is not to show the emergence of ferromagnetism in 2D materials per se, but to provide an estimate of the $T_C$, which this model does effectively. We want to note that the 2D Ising model has been used to extract $T_C$ for other 2D ferromagnets in recent literature.\cite{Miao} }

\begin{figure*}
    \centering
    \includegraphics[width=1.6\columnwidth]{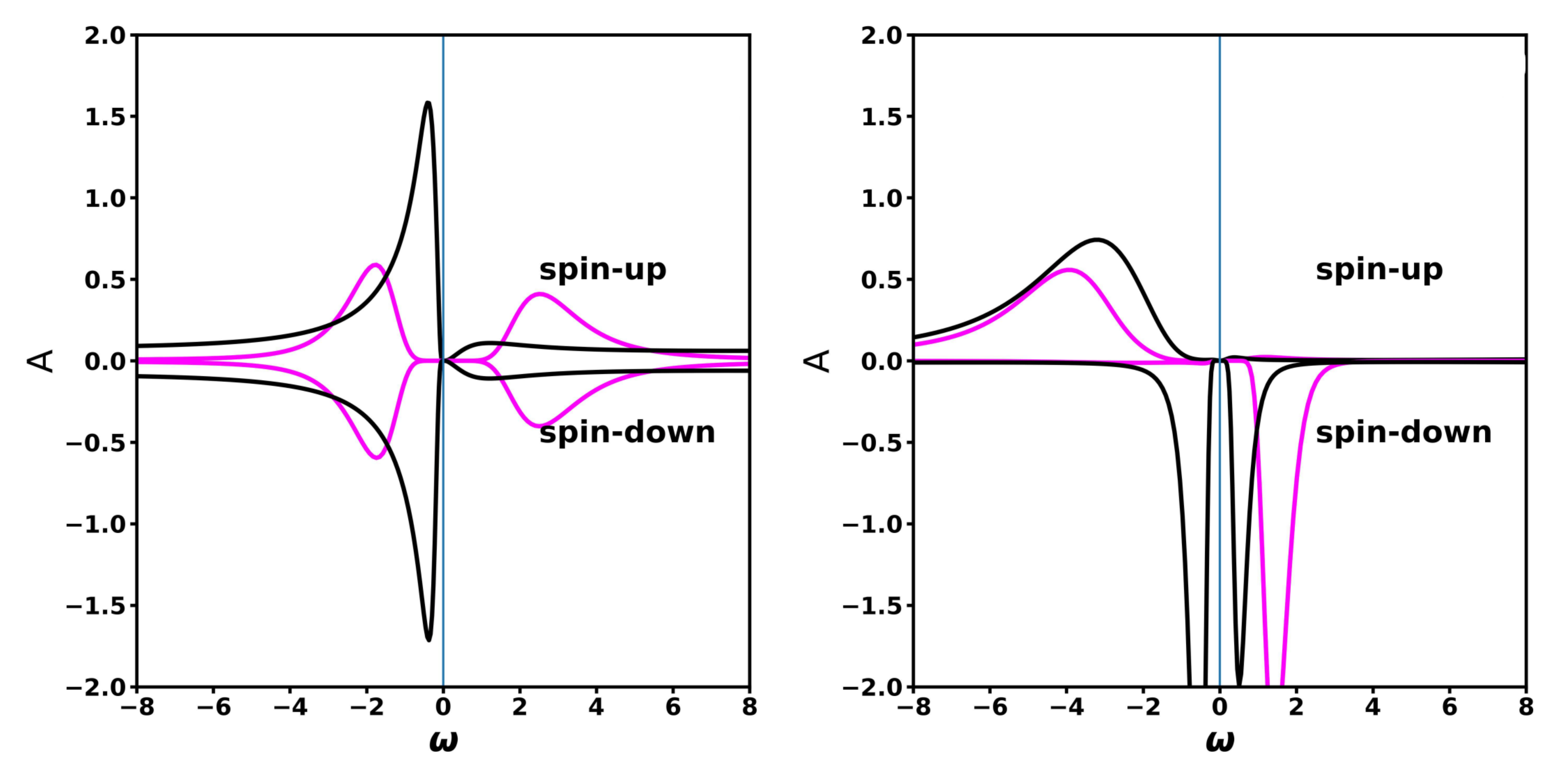}
    \caption{(Color online) DMFT correlated spectral functions for $U=3.5$\,eV and $J_H=0.5$\,eV. Left: Paramagnetic solution at \textcolor{black}{inverse temperature} $\beta=40$\,eV$^{-1}$. Right: Spin-polarized solution at \textcolor{black}{inverse temperature} $\beta=200$\,eV$^{-1}$. \textcolor{black}{The black curves represent the sum of the three almost degenerate orbitals (one doublet+one singlet) at lower energy, and the magenta curve is the sum of two higher energy degenerate orbitals (doublet).} The spectra have been obtained using the 
    maximum entropy method of analytic continuation. }
    \label{fig3}
\end{figure*}

\subsection{DMFT calculations}
\begin{figure}
    \centering
    \includegraphics[width=\columnwidth]{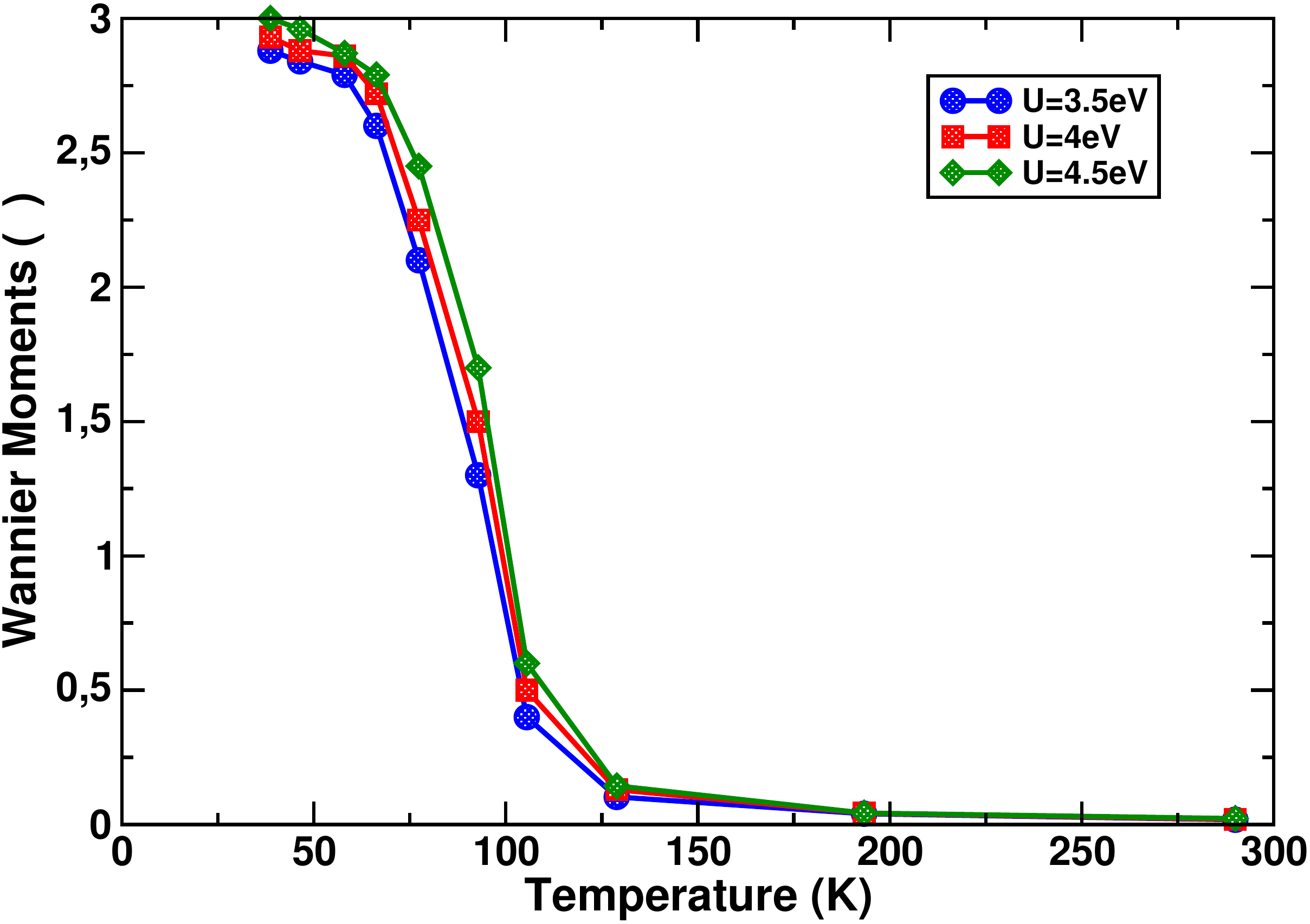}
    \caption{(Color online) Wannier magnetic moment versus temperature obtained from a DMFT calculation for \textcolor{black}{$U=3.5$, 4, and 4.5\,eV}, and $J_H=0.5$\,eV.}
    \label{fig4}
\end{figure}

We carry out DFT+DMFT calculations to include electronic correlations more appropriately, and try to estimate the Curie temperature for the paramagnetic to ferromagnetic transition. Our paramagnetic DFT band structure calculations using \textsc{wien2k} reveal a metallic solution, with \textcolor{black}{lower-energy doublet+singlet $d$} orbitals \textcolor{black}{at roughly $-0.8$\,eV} below Fermi energy, and \textcolor{black}{higher-energy doublet} orbitals crossing the Fermi energy, \textcolor{black}{in exact agreement with} our VASP calculations shown in Fig.~\ref{fig2}.

We first carry out paramagnetic DMFT calculations at \textcolor{black}{inverse temperature} $\beta=40$\,eV$^{-1}$, including \textcolor{black}{all five Co $d$ orbitals to allow for high spin solutions. Further explanation on the Wannierization is provided in Appendix A}. The correlated spectral function for $U=3.5$\,eV and $J_H=0.5$\,eV is shown in Fig.~\ref{fig3}. We see an insulating solution with a very small band gap at the point of a metal-to-insulator transition. \textcolor{black}{Within DMFT,} the \textcolor{black}{lower energy singlet+doublet} orbitals are seen to be majorly occupied with a \textcolor{black}{total occupancy of $n=5$}. The \textcolor{black}{higher-energy doublet} on the other hand is occupied by two electrons and is, thus, half filled. \textcolor{black}{These different fillings result in very different response to electron interactions, and lead to strongly orbital-selective correlations}. We did calculations also for increased 
$U=4$ and 4.5\,eV, using the same $J_H=0.5$\,eV (not shown). The gap within the \textcolor{black}{higher energy doublet} orbitals increases with increasing $U$, but there is almost no change in the gap within the \textcolor{black}{lower-energy singlet+doublet} orbitals. This is due to the very different reaction of multiorbital problems as function of their occupation.\cite{georgesHund} \textcolor{black}{As a result, the half-filled higher-energy doublet shows strong correlations, and dependence on $U$, whereas the small gap of the lower-energy singlet+doublet states does not vary much as function of $U$. We want to note here that the DMFT solution shows the polarised occupations-five electrons in the singlet+doublet states and two electrons in the higher-energy doublet-necessary for a high-spin magnetic solution already in the paramagnetic state.}

\textcolor{black}{We carried out DMFT calculations at lower value of $U=1.3$\,eV, $J=0.3$\,eV, which clearly shows a metallic solution in the higher energy doublet.
%but there is almost not change in the pseudogap like feature in the lower energy singlet+doublet orbitals. 
This is shown in Appendix B in Fig.~\ref{figB1}. }
%The vicinity to the metal insulator transition is due to the fact that throughout the entire range of U=3.5 to 4.5 there is always a small pseudo gap like feature, due to the spectral function of the lower energy singlet+doublet not changing significantly with U, due to the weakly correlated nature owing to the 5 electrons in its 6 orbitals however the higher energy doublet being half filled reacts strongly to Hubbard U as expected.}

In the next step we investigate the spin-polarisation in the DMFT solutions. Starting from the paramagnetic solutions, we introduce a spin splitting in the real part of the self energies, and let the DMFT iterative cycle converge to a possibly spin-split solution. We carry out the calculations at various values of inverse temperature $\beta$ between 40 and 250\,eV$^{-1}$. 

 At $\beta=40$\,eV$^{-1}$, the calculation converges still to a paramagnetic state, but when reducing the temperature we find a transition to a ferromagnetic ground state. The spectral function at $\beta=200$\,eV$^{-1}$ is shown in Fig.~\ref{fig3} for $U=3.5$\,eV. Again, a very similar variation is seen in the electronic structure with changes in $U$ values. We see a clear splitting between the up and down spin channels, and a band gap of 0.2\,eV, slightly larger than in the paramagnetic phase. 
 %Here we observe the correct distribution of electrons between the \textcolor{black}{lower energy singlet+doublet} and \textcolor{black}{higher energy doublet} states. 
 We observe that the spin-up channel for \textcolor{black}{ the higher-energy doublet} orbitals is occupied with two electrons, while the spin-down channel for the same is empty. For the \textcolor{black}{lower-energy singlet+doublet} orbitals, the spin-up channel is fully filled while the spin-down channel is only partially filled with two electrons. This gives the total magnetic moment of 3\,$\mu_B$ coming from two unpaired electrons in the \textcolor{black}{higher-energy doublet} orbitals and one unpaired electron in the \textcolor{black}{almost degenerate lower-energy singlet+doublet} orbitals. 
 %We get the same conclusion from the analysis of the occupancies from the converged density matrix.

Next, we look at the temperature dependence of the ferromagnetic solution, as we wish to determine the Curie temperature from a DMFT perspective. We plot the Wannier magnetic moments on Co, obtained from the density matrix of the spin-split DMFT solution, in Fig.~\ref{fig4}. It is obvious that a transition from a paramagnetic to a ferromagnetic state occurs at around $\beta=125$\,eV$^{-1}$, which corresponds roughly to a temperature of 100\,K. It is interesting to note that here as well changing the value of $U$ from 3.5 to 4 to 4.5 does not change this value of the transition temperature \textcolor{black}{much}, unlike in the case of the DFT+U studies. This can be correlated to the fact that CoBr$_2$ is an insulator with a small band gap in the paramagnetic phase in a quite large range of interaction values. \textcolor{black}{As already discussed above}, for this special case of half-filled \textcolor{black}{higher-energy doublet} and $5/6$-filled \textcolor{black}{lower-energy singlet+doublet} orbitals, the total band gap does not significantly change with $U$. As a result, the system lies quite robustly at the phase transition point between an insulator and a metal, which has been shown to be a hot spot for large magnetic transition temperatures.\cite{onebanddmft,jernej,alen2017} 

\textcolor{black}{However, it should also be noted that DMFT, too, overestimates the magnetic transition temperatures. For 3D bulk systems, this overestimation is normally between a factor close to 1 up to a factor of 2,\cite{PhysRevB.86.035152,jernej} depending on the system under investigation. In layered systems, this factor can be even larger, since finite-wave-length spin fluctuations are stronger. For instance, it has been found in technetium oxides that the 3D variant SrTcO$_3$ has a transition temperature of roughly 1000\,K,\cite{jernej} whereas the layered variant Sr$_2$TcO$_4$ was predicted to have a transition temperature of around 550\,K.\cite{alen2017} Single-site DMFT would rather give similar estimates for the $T_C$ in the two cases. Given all the uncertainities, we can estimate a $T_C$ in the range of 30 to 50\,K, which is more in line with the prediction in Ref.~\onlinecite{Liu2018} than with the small $T_C$'s in other previous studies.\cite{Botana2019} }
%we can estimate at least a Curie Temperature of $\sim$ 50K from the DMFT calculations, which is more in line with Ref.~\onlinecite{Liu2018}.

\section{Conclusion}

 We have investigated the influence and importance of electronic correlations for a monolayer of CoBr$_2$. 
 This system can easily be obtained from the bulk van-der-Waals crystal by exfoliation. 
 %
 %This system can be easily obtained by exfoliation  which is a member shown in case of monolayers of MX$_2$ type van der Waals crystals, taking the example of CoBr$_2$ monolayer which is a dynamically stable and easily exfoliable material, that correlations play an important role in the overall electronic structure and particularly in the estimation of Curie temperature $T_C$ for the paramagnetic to ferromagnetic transition. 
 First, we have applied a standard methodology for the estimation of the transition temperature, which is a combination of DFT+U for the calculation of exchange couplings, and a subsequent solution of a classical spin model using Monte Carlo techniques. We find that the transition temperature varies substantially with the interaction parameter $U$ that is used in the DFT+U treatment. Nevertheless, $T_C$'s in the range of $60$ to $140$\,K can be obtained for reasonable values of $U$.
 
 Treating correlations within DMFT leads on first sight to similar transition temperatures. However, the physical picture is slightly different. Different to DFT+U calculations, we see only a marginal dependence of the single-particle gap on the Hubbard parameter $U$. As a result, the system is placed very robustly at the vicinity of a metal-to-insulator transition. This point in the phase diagram has been shown to be very beneficial for magnetic properties, as we also see here. A careful estimation of the transition temperature, also taking into account possible over-estimations due to the mean-field nature of the theories \textcolor{black}{and the dimensionality of the problem, gives a range of $T_c\sim 30$ to $\sim 50$\,K}. This is a quite remarkable transition temperature for a 2D material. Furthermore, the concept to find materials in the vicinity of metal-insulator phase transitions to find good magnets is corroborated by this study, and might be exploited in the future to enhance even more the Curie temperatures of these layered materials.

 %tudies that in case of DFT+U, the value of magnetic superexchange depends significantly on the choice of Hubbard U, which in turn controls the transition temperature obtained from Monte Carlo solutions of Ising model based on the lattice structure, over and above the usual overestimation of Ising type models in such cases. We show that DMFT is a useful tool in this case to determine primarily the electronic structure of materials at the verge of a Mott transition. Estimation of the Curie temperatures from DMFT is more robust with respect to Hubbard and Coulomb parameters since it does not have the U dependence of $T_C$ and the Curie temperature obtained albeit overestimated is independent of the choice of Hubbard U, due to the fact that the paramagnetic spectra shows the material to be at the point of a metal insulator transition. It is to be noted that a static GGA+U method overestimates the band gap and an ad hoc inclusion of Hubbard U is not sufficient without a proper description of the quasiparticle peaks which is taken care of in DMFT calculations.

\appendix
\section{Wannier Projections}
\begin{figure}
    \centering
    \includegraphics[width=\columnwidth]{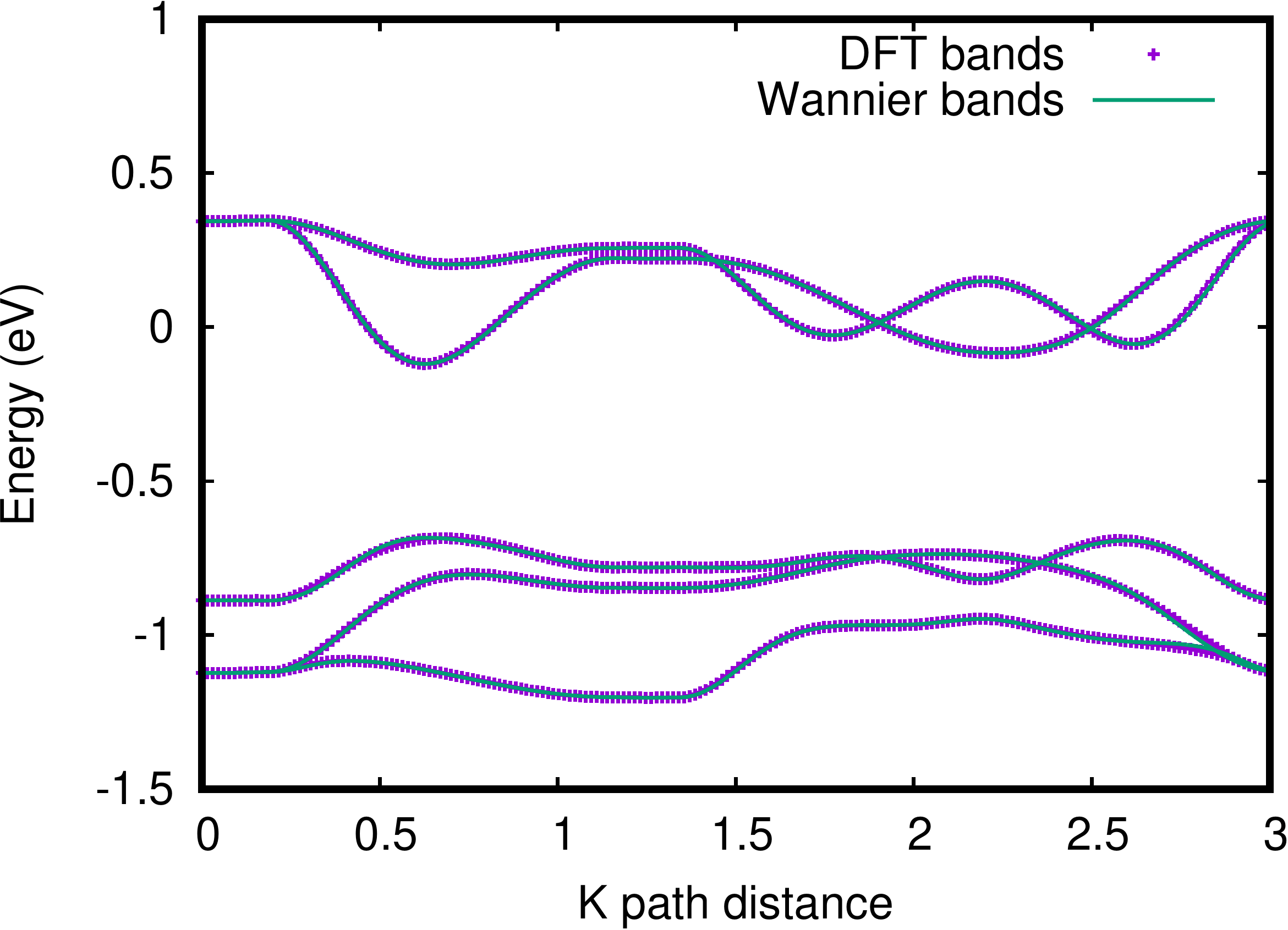}
    \caption{\textcolor{black}{(Color online) The non-magnetic DFT band structure superimposed with the effective Wannier projected bands along a typical path through the Brillouin zone.}}
    \label{figA1}
\end{figure}
Here we show in Fig.~\ref{figA1} the Wannier projected $d$ bands superposed on the \textcolor{black}{non-magnetic DFT band structure obtained from \textsc{Wien2k}}, which we consider for setting up the DMFT calculations. \textcolor{black}{The overall structure is that there are three bands at binding energies between -1.2 and -0.7\,eV, and two degenerate bands around the Fermi level. It can be easily seen that the three lower-energy bands are further split up into a singlet and a doublet. The calculation of the orbital energies from the local Wannier Hamiltonian shows that the singlet has orbital energy of $-0.849$\,eV, whereas the doublet is located at the almost degenerate energy $-0.842$\,eV. The two higher energy states-the doublet-are located at 0.043\,eV. That the three lower-energy states are almost degenerate can also be seen from the local density matrix, which give orbital occupations of $n=1.953$ for the singlet and the doublet, resp. For the states at the Fermi level we get orbital occupancies of $n=0.57$.} 

\textcolor{black}{Looking at the band structure in Fig.~\ref{figA1}, one could be tempted to construct Wannier orbitals only for the two bands around the Fermi level. This procedure, however, would result in a complete filling of the lower-energy singlet+doublet states with $n=6$, leading to $n=1$ in the effective two-band Wannier Hamiltonian.
This in turn allows only for low-spin state solutions with 1$\mu_B$ instead of the expected high spin state of 3$\mu_B$. Therefore, to allow for the high spin state all five d bands are considered in the calculation. In this 5-band calculations, the total filling of the Wannier orbitals is always $n=7$. The hybridisation with the Br $p$ states is as usual taken care of by the Wannier construction and orthonormalisation.}

%This causes a re-arrangement in the fillings with 5 electrons in the 3 low lying bands making them weakly correlated and 2 electrons in the higher energy bands making them half filled and strongly correlated, further supporting our findings. 

\section{Paramagnetic DMFT at smaller Hubbard $U$}
\begin{figure}
    \centering
    \includegraphics[width=\columnwidth]{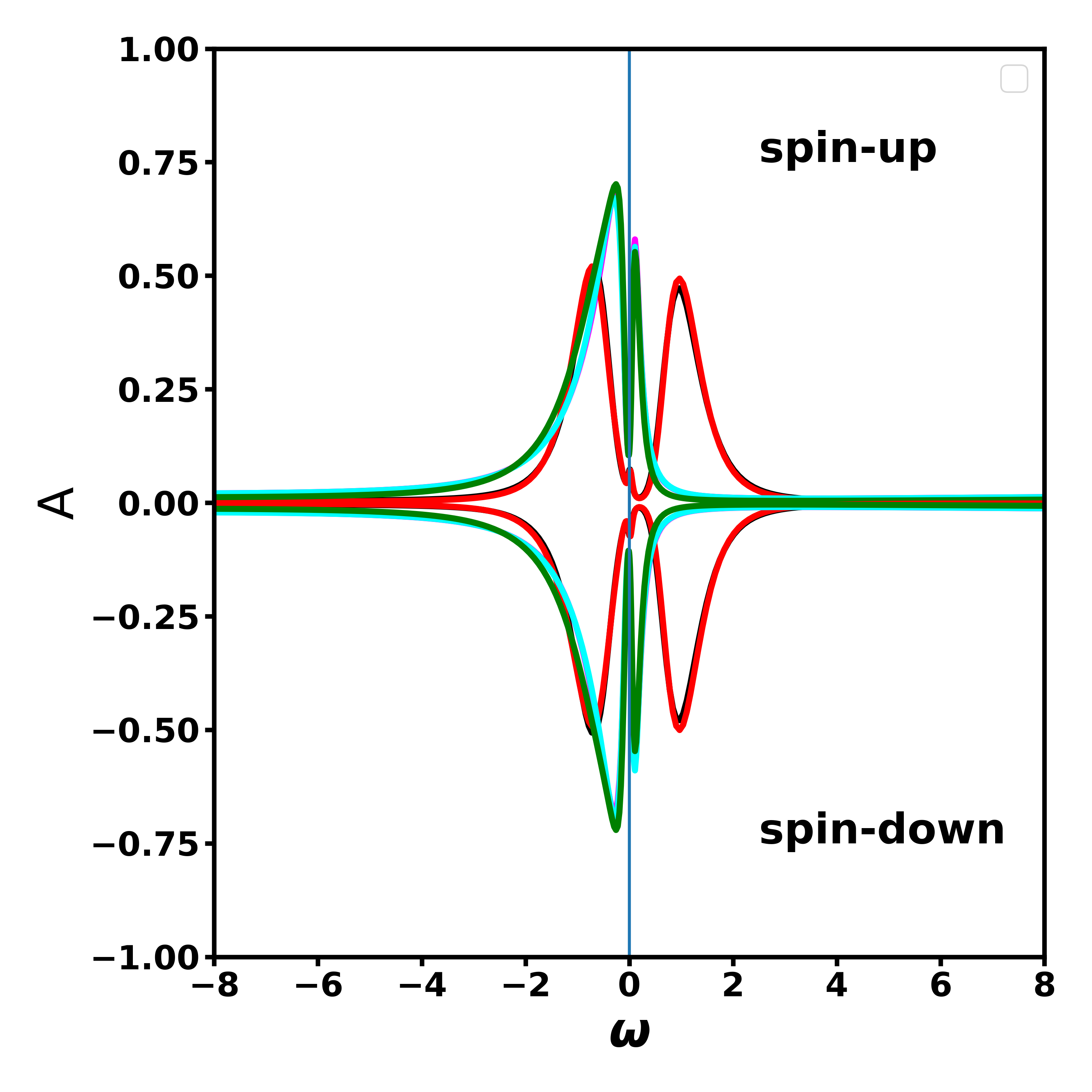}
    \caption{\textcolor{black}{(Color online) The paramagnetic DMFT correlated spectral functions for $U=1.5$\,eV and $J_H=0.3$\,eV at inverse temperature $\beta=40$\,eV$^{-1}$. Green, magenta and cyan curves represent the three lower energy orbitals which are as seen here almost degenerate, while the red and the black represent the two higher energy orbitals which are also seen to be degenerate.}}
    \label{figB1}
\end{figure}
\textcolor{black}{As discussed in the main text, the overall spectral gap is rather independent of the Hubbard $U$ at reasonable interaction values. Here, we show the paramagnetic correlated spectral function for a smaller Hubbard interaction value $U=1.5$\,eV. In order to keep the interaction values in the Kanamori Hamiltonian physically meaningful, we also decreased the Hund's coupling to $J_H=0.3$\,eV, such that $U-3J_H$ remains positive. From Fig.~\ref{figB1} one can see that the gap in the higher-energy doublet has closed, and a small quasi-particle feature has emerged at the Fermi level, leading to a metallic state. Also the distribution of spectral weight in the other orbitals has changed, and significantly shifted towards the Fermi level. The gap that has been very clear in Fig.~\ref{fig3} has become at most a pseudo gap.} 
%lower-lying orbitals has shifted to smaller energies around Fermi. Due to the metallic solution, the orbital occupancies have changed, in the higher-energy orbitals
%from $n=2$ to a slightly non-integer value of 2.092. Eventually, for $U$ and $J_H$ approaching zero, one has to reproduce the above mentioned occupancies in the non-interacting Wannier Hamiltonian.}
%Incidentally it is to be noted that the three low lying almost degenerate orbitals, with 5 electrons are very weakly correlated and is almost unaffected by the change in Hubbard U while the two higher energy half filled orbitals are strongly correlated and becomes metallic on reduction of Hubbard U, confirming the presence of a metal insulator transition in the system.}

\begin{acknowledgments}
This work has been funded by the Austrian Science Fund (FWF), START project Y746. Calculations have partly been performed on the dcluster of TU Graz.
\end{acknowledgments}

\bibliography{main}

\end{document}